\documentstyle[12pt,epsfig]{article}

\newcommand{\be}{\begin{equation}}
\newcommand{\ee}{\end{equation}}
\newcommand{\Dlt}{\Delta}
\newcommand{\dlt}{\delta}

\newcommand{\br}{{\bf r}}

\newcommand{\bk}{{\bf k}}

\newcommand{\bP}{{\bf P}}

\newcommand{\vp}{\varphi}
\newcommand{\ep}{\varepsilon}

\newcommand{\ra}{\rightarrow}
\newcommand{\sgm}{\sigma}

\newcommand{\gm}{\gamma}
\newcommand{\om}{\omega}

\newcommand{\dgr}{\dagger}

\begin{document}

\begin{center}
{\Large{\bf Condensate and superfluid fractions for varying interactions 
and temperature} \\ [5mm]
V.I. Yukalov$^1$ and E.P. Yukalova $^2$} \\ [3mm]

{\it $^1$Bogolubov Laboratory of Theoretical Physics, \\
Joint Institute for Nuclear Research, Dubna 141980, Russia \\ [3mm]

$^2$Department of Computational Physics, Laboratory of Information 
Technologies, \\
Joint Institute for Nuclear Research, Dubna 141980, Russia}

\end{center}

\vskip 2cm

\begin{abstract}

A system with Bose-Einstein condensate is considered in the frame of the 
self-consistent mean-field approximation, which is conserving, gapless, and 
applicable for arbitrary interaction strengths and temperatures. The main 
attention is paid to the thorough analysis of the condensate and superfluid 
fractions in the whole range of the interaction strength, between zero and 
infinity, and for all temperatures between zero and the critical point $T_c$. 
The normal and the anomalous averages are shown to be of the same order for 
almost all interactions and temperatures, except the close vicinity of $T_c$. 
But even in the vicinity of the critical temperature, the anomalous average 
cannot be neglected, since only in the presence of the latter the phase 
transition at $T_c$ becomes of second order, as it should be. Increasing 
temperature influences the condensate and superfluid fractions in a similar 
way, by diminishing them. But their behavior with respect to the interaction 
strength is very different. For all temperatures, the superfluid fraction 
is larger than the condensate fraction. These coincide only at $T_c$ or under 
zero interactions. For asymptotically strong interactions, the condensate is 
almost completely depleted, even at low temperatures, while the superfluid 
fraction can be close to one.

\end{abstract}

\vskip 1cm

{\bf PACS number(s)}: 03.75.Hh, 03.75.Kk, 03.75.Nt, 05.30.Ch

\newpage

\section{Introduction}

The relation between Bose-Einstein condensation (BEC) and superfluidity 
is of long-standing interest. The most thorough studies, both experimental 
and theoretical, of Bose-system properties have been accomplished for the 
region of low temperatures and weak interactions (see review works [1--7]), 
where the Bogolubov approximation [8,9] is applicable. In this region, 
almost the entire system is Bose-condensed, being just slightly depleted 
by interactions, at the same time, practically the whole system is 
superfluid. When the condensate and superfluid fractions are so close 
to each other, one often uses as synonyms the terms "Bose-condensed" and 
"superfluid".

The opposite situation occurs for strongly interacting liquids, such 
as superfluid $^4$He, where the superfluid fraction at low temperatures 
almost reaches one, while the condensate fraction never exceeds the 
values of the order of $10\%$. The superfluid properties of helium are 
among the best-measured in experimental physics, as can be inferred, e.g., 
from the books [10--13]. The BEC fraction in superfluid helium, has been 
measured by using the $x$-ray scattering [14] and deep-inelastic neutron 
scattering [15]. Theoretical investigation, because of strong interactions 
between helium atoms, is rather complicated and mainly is done numerically, 
for instance, by means of Monte Carlo techniques [16,17].

It would be important to understand the behavior of the condensate and 
superfluid fractions of the same system in the whole region of varying 
interaction strength and temperature. Present-day Feshbach-resonance 
techniques do allow for the variation of the interaction strength in a 
very wide range [18,19]. It is the aim of the present paper to investigate 
the condensate and superfluid fractions for all temperatures between zero 
and the critical temperature $T_c$ and for all interaction strengths 
between zero and infinity. Such an analysis can be accomplished in the 
frame of the self-consistent mean-field theory [20--25], which is conserving, 
gapless, satisfies all thermodynamic relations and conservation laws. It was 
shown that this theory yields good agreement with Monte Carlo simulations for 
weak as well as strong interactions [24] and can also be applied for Bose 
systems in random potentials with arbitrary strong strength of disorder [25].

Studying here the condensate and superfluid fractions, we analyse their 
properties both analytically and numerically. We use the system of units 
with $\hbar\equiv 1$ and $k_B\equiv 1$.

\section{Uniform Equilibrium System}

Let us consider a uniform equilibrium Bose system with BEC. The 
appearance of BEC, as is known [26--28], is equivalent to the gauge 
symmetry breaking. The most convenient way to realize the latter is 
by means of the Bogolubov operator shift [29,30] representing the Bose 
field operator as the sum

\be
\label{1}
\hat\psi(\br) \equiv \eta(\br) +\psi_1(\br) \; .
\ee
The first term here is the condensate wave function normalized to the 
number of condensed atoms
\be
\label{2}
N_0 = \int |\eta(\br)|^2 \; d\br \; ,
\ee
and the second term is the field operator of uncondensed atoms, 
satisfying the Bose commutation relations and the normalization to the 
number of uncondensed atoms
\be
\label{3}
N_1 = \int <\psi_1^\dgr(\br) \psi_1(\br)> d\br \; .
\ee
The angle brackets mean, as usual, statistical averaging. The condensate 
function $\eta(\br)$ and the operator of uncondensed atoms $\psi_1(\br)$ 
are treated as independent variables, orthogonal to each other,
\be
\label{4}
\int \eta^*(\br) \psi_1(\br)\; d\br = 0 \; .
\ee
The total average density is the sum
\be
\label{5}
\rho = \rho_0 + \rho_1 
\ee
of the condensate density $\rho_0$ and the density $\rho_1$ of 
uncondensed atoms,
\be
\label{6}
\rho_0 \equiv \frac{N_0}{V} \; , \qquad
\rho_1 \equiv \frac{N_1}{V} \; ,
\ee
where $V$ is the system volume. The atoms are assumed to interact with 
each other through the local interaction potential
\be
\label{7}
\Phi(\br) = \Phi_0 \dlt(\br) \; , \qquad \Phi_0 \equiv 
4\pi\; \frac{a_s}{m} \; ,
\ee
with $a_s$ being the scattering length and $m$ being atomic mass.

For a uniform equilibrium system, the condensate function is constant,
\be
\label{8}
\eta(\br) =\sqrt{\rho_0} \; .
\ee
The field operator of uncondensed atoms can be expanded over plane waves
$$
\vp_k(\br) \equiv \frac{e^{i\bk\cdot\br}}{\sqrt{V}} \; , 
$$
which gives
\be
\label{9}
\psi_1(\br) = \sum_k a_k \vp_k(\br) \; .
\ee
From the orthogonality equation (4) it follows that
\be
\label{10}
\lim_{k\ra 0} a_k = 0 \; .
\ee
Therefore, in expansion (9), the summation is over $\bk\neq 0$. This 
condition can either be shown explicitly in Eq. (9) or just one can keep 
in mind property (10).

Accomplishing the Bogolubov shift for the field operator (1), we have the 
grand Hamiltonian
\be
\label{11}
H = \sum_{n=0}^4 H^{(n)}
\ee
consisting of five terms, labelled according to their order with respect 
to $a_k$ and $a_k^\dgr$. The zero-order term does not contain the field 
operators of uncondensed atoms
\be
\label{12}
H^{(0)} = \left ( \frac{1}{2}\; \rho_0 \Phi_0 - \mu_0 
\right ) N_0 \; .
\ee
The first-order term is identically zero,
\be
\label{13}
H^{(1)} = 0 \; ,
\ee
because of the orthogonality equation (4). The second-order term is
\be
\label{14}
H^{(2)} = \sum_k \left [ \left ( \frac{k^2}{2m} +
2\rho_0 \Phi_0 - \mu_1 \right ) a_k^\dgr a_k + 
\frac{1}{2}\; \rho_0 \Phi_0 \left ( a_k^\dgr a_{-k}^\dgr + 
a_{-k} a_k \right ) \right ] \; .
\ee
Respectively, one has the third-order term
\be
\label{15}
H^{(3)} = \sqrt{ \frac{\rho_0}{V} } \; \Phi_0 \; 
\sum_{p,q} \left ( a_q^\dgr a_{q-p} a_p + 
a_p^\dgr a_{q-p}^\dgr a_q \right )
\ee
and the fourth-order term
\be
\label{16}
H^{(4)} = \frac{\Phi_0}{2V} \; \sum_{k,p,q} a_p^\dgr a_q^\dgr
a_{k+p} a_{q-k} \; .
\ee
In Eqs. (14), (15), and (16), the sums do not contain the terms with 
the operators $a_k$, for which $\bk=0$, due to the limiting condition 
(10).

The momentum distribution of uncondensed atoms is given by the normal 
average
\be
\label{17}
n_k \; \equiv \; < a_k^\dgr a_k > \; .
\ee
Because of the broken gauge symmetry, there also appears the anomalous 
average
\be
\label{18}
\sgm_k \; \equiv \; < a_k a_{-k} > \; .
\ee
The normal and anomalous averages are equally important and neither 
of them can be neglected. Omitting the anomalous average would make 
the theory not self-consistent and would yield a spurious system 
instability [5,31]. Summing Eqs. (17) and (18) gives the density of 
uncondensed atoms
\be
\label{19}
\rho_1 = \frac{1}{V} \; \sum_k n_k
\ee
and the anomalous average
\be
\label{20}
\sgm_1 = \frac{1}{V} \; \sum_k \sgm_k \; ,
\ee
for which the value $|\sgm_1|$ defines the density of pair-correlated 
atoms [32].

Applying for terms (15) and (16) the Hartree-Fock-Bogolubov approximation 
and involving the Bogolubov canonical transformation
$$
a_k = u_k b_k + u_{-k}^* b_{-k}^\dgr \; ,
$$
we reduce [24] the grand Hamiltonian (11) to the form
\be
\label{21}
H = E_B + \sum_k \ep_k b_k^\dgr b_k \; ,
\ee
in which
\be
\label{22}
E_B = -\left ( \frac{1}{2}\; \rho_0 + 2\rho_1 +\sgm_1 
\right ) \Phi_0 N_0 \; - \; \frac{\Phi_0}{2\rho} \left (
2\rho_1^2 + \sgm_1^2 \right ) N + \frac{1}{2} \; 
\sum_k (\ep_k - \om_k)
\ee
is a nonoperator quantity;
\be
\label{23}
\ep_k =\sqrt{ (ck)^2 +\left ( \frac{k^2}{2m}\right )^2 }
\ee
is the Bogolubov-type spectrum, but with the sound velocity $c$ defined 
by the equation
\be
\label{24}
mc^2 =  ( \rho_0 +\sgm_1 ) \Phi_0 \; .
\ee

For the momentum distribution (17), we get
\be
\label{25}
n_k = \frac{\om_k}{2\ep_k} \; {\rm coth} \left (
\frac{\ep_k}{2T} \right ) \; - \; \frac{1}{2} \; ,
\ee
where $T$ is temperature and
\be
\label{26}
\om_k \equiv \frac{k^2}{2m} + mc^2 \; .
\ee
And for the anomalous average (18), we find
\be
\label{27}
\sgm_k = -\; \frac{mc^2}{2\ep_k} \; {\rm coth}\left (
\frac{\ep_k}{2T} \right ) \; .
\ee
We may notice that
$$
n_k + \frac{\om_k}{mc^2}\; \sgm_k + \frac{1}{2} = 0 \; ,
$$
which indicates that $\sgm_k$ is, generally, of the same order as 
$n_k$.

Using Eq. (25) and passing in the standard way from summation over 
momenta to their integration, we obtain for the density of uncondensed 
atoms (19)
\be
\label{28}
\rho_1 = \frac{(mc)^3}{3\pi^2} \left \{ 1 + 
\frac{3}{2\sqrt{2}} \; \int_0^\infty \; \left (
\sqrt{1+x^2} - 1\right )^{1/2} \left [ {\rm coth}\left ( 
\frac{mc^2}{2T}\; x\right ) - 1 \right ] \; dx \right \} \; .
\ee
For the anomalous average (20), using Eq. (27), we find
\be
\label{29}
\sgm_1 = \sgm_0 \; - \; \frac{(mc)^3}{2\sqrt{2}\; \pi^2} \;
\int_0^\infty \; 
\frac{\left ( \sqrt{1+x^2}-1\right )^{1/2}}{\sqrt{1+x^2}}
\left [ {\rm coth}\left ( \frac{mc^2}{2T}\; x \right ) -1
\right ] \; dx \; ,
\ee
where in the calculation of the term
\be
\label{30}
\sgm_0 = \frac{(mc)^2}{\pi^2}\; \sqrt{m \rho_0 \Phi_0 }
\ee
the dimensional regularization [24] is employed, in line with the 
general rules of the dimensional regularization [5,33].

\section{Condensate and Superfluid Fractions}

Our main aim here is to study the condensate and superfluid fractions. 
The condensate fraction
\be
\label{31}
n_0 \equiv \frac{N_0}{N} = 1  - n_1
\ee
can be found by calculating the normal fraction
\be
\label{32}
n_1 \equiv \frac{N_1}{N} = \frac{\rho_1}{\rho} \; ,
\ee
with the normal density (28).

The superfluid fraction can be represented [2,25] as
\be
\label{33}
n_s = 1 - \frac{2Q}{3T} \; ,
\ee
where $Q$ is the dissipated heat,
\be
\label{34}
Q \equiv \frac{\Dlt^2(\hat\bP)}{2mN} \; ,
\ee
expressed through the dispersion
$$
\Dlt^2(\hat\bP) \; \equiv \; < \hat\bP^2> - <\hat\bP>^2
$$
of the total momentum operator $\hat\bP$. For the dissipated heat, in 
the considered mean-field approximation, we have
\be
\label{35}
Q = \frac{1}{\rho} \; \int \frac{k^2}{2m} \left ( n_k + n_k^2 - 
\sgm_k^2 \right ) \frac{d\bk}{(2\pi)^3} \; .
\ee
Substituting here Eqs. (25) and (27), we get
\be
\label{36}
Q = \frac{1}{8m\rho} \; \int \; \frac{k^2}{{\rm sinh}^2(\ep_k/2T)} \;
\frac{d\bk}{(2\pi)^3} \; .
\ee
The latter equation can be transformed to
\be
\label{37}
Q = \frac{(mc)^5}{\sqrt{2}(2\pi)^3 m\rho} \; \int_0^\infty \; 
\frac{(\sqrt{1+x^2}-1)^{3/2} \;x\; dx}
{\sqrt{1+x^2}\;{\rm sinh}^2(mc^2x/2T)} \; .
\ee

For the following analysis, it is convenient to introduce dimensionless 
quantities. We define the gas parameter
\be
\label{38}
\gm \equiv \rho^{1/3} a_s \; ,
\ee
measuring the interaction strength, and the dimensionless temperature
\be
\label{39}
t \equiv \frac{mT}{\rho^{2/3}} \; .
\ee
We introduce the dimensionless sound velocity
\be
\label{40}
s \equiv \frac{mc}{\rho^{1/3}}
\ee
and the dimensionless anomalous average
\be
\label{41}
\sgm \equiv \frac{\sgm_1}{\rho} \; .
\ee

In terms of these notations, the dimensionless velocity (40), in view 
of Eq. (24), satisfies the equation
\be
\label{42}
s^2 = 4\pi \gm (n_0 + \sgm ) \; .
\ee
The condensate fraction (31) is expressed through the normal fraction 
(32), for which we have
\be
\label{43}
n_1 = \frac{s^3}{3\pi^2} \left\{ 1 + \frac{3}{2\sqrt{2}} \; 
\int_0^\infty \; \left ( \sqrt{1+x^2} -1 \right )^{1/2}
\left [ {\rm coth}\left ( \frac{s^2 x}{2t}\right ) -1 \right ] \;
dx \right \} \; .
\ee
The anomalous average (41), according to Eqs. (29) and (30), becomes
\be
\label{44}
\sgm = \frac{2s^2}{\pi^{3/2}}\; \sqrt{\gm n_0} \; - \;
\frac{s^3}{2\sqrt{2}\; \pi^2} \; \int_0^\infty \; 
\frac{(\sqrt{1+x^2}-1)^{1/2}}{\sqrt{1+x^2}} \; \left [
{\rm coth}\left ( \frac{s^2x}{2t}\right ) - 1 \right ]\; dx \; .
\ee
And the superfluid fraction (33) takes the form
\be
\label{45}
n_s =  1\; - \; \frac{s^5}{6\sqrt{2}\; \pi^2 t} \; 
\int_0^\infty \; \frac{(\sqrt{1+x^2}-1)^{3/2} x\; dx}
{\sqrt{1+x^2}\;{\rm sinh}^2(s^2x/2t)} \; .
\ee
Equations (42) to (45), together with the relation $n_0=1-n_1$, define 
all characteristics we wish to investigate.

\section{Varying Interactions and Temperature}

Our aim is to study Eqs. (42) to (45) for the varying interaction 
strength $\gm\geq 0$ and temperature $t\geq 0$. First, we find 
analytic expressions for low temperatures and for the temperature 
close to the critical point.

\vskip 5mm

{\large{\bf A. Low Temperature}}

\vskip 3mm

At sufficiently low temperature, such that
\be
\label{46}
\frac{t}{s^2} \ll 1 \; ,
\ee
we find from Eqs. (43) to (45) the asymptotic expansions for the 
normal fraction
\be
\label{47}
n_1 \simeq \frac{s^3}{3\pi^2} + \frac{t^2}{12s} \; ,
\ee
the anomalous average
\be
\label{48}
\sgm \simeq \frac{2s^2}{\pi^{3/2}}\; \sqrt{\gm n_0} \; - \;
\frac{t^2}{12s} \; ,
\ee
and for the superfluid fraction
\be
\label{49}
n_s \simeq 1 \; - \; \frac{2\pi^2 t^4}{45 s^5} \; .
\ee
We may notice that between the condensate and superfluid fractions 
there is the relation
\be
\label{50}
( 1 - n_s) (1 - n_0)^{5/3} \simeq 
\frac{2t^4}{135\pi(9\pi)^{1/3}} \; .
\ee
Substituting these expansions into Eq. (42), we obtain the asymptotic 
behavior of the dimensionless sound velocity
\be
\label{51}
s \simeq s_0 + at^2 \qquad (t \ra 0) \; .
\ee
Here the zero-temperature term is defined by the equation
\be
\label{52}
\pi \left ( 4 \gm s_0^3 + 3\pi s_0^2 - 12 \pi^2 \gm 
\right )^2 = 192 \gm^3 s_0^4 \left ( 3\pi^2 - s_0^3 \right ) \; .
\ee
The coefficient $a$ in the second term of Eq. (51) is
\be
\label{53}
a = \frac{3\pi^2(4\gm s_0^3-\pi s_0^2-12\pi^2\gm)}
{16(2\gm s_0^6 + 3\pi s_0^5-18\pi^2\gm s_0^3+36\pi^4\gm)} \; .
\ee
Using expansion (51) in Eq. (48) gives the anomalous average
\be
\label{54}
\sgm \simeq \sgm_0 + bt^2 \; ,
\ee
in which
\be
\label{55}
\sgm_0 = \frac{4\gm s_0^3+3\pi s_0^2 - 12\pi^2\gm}{12\pi^2\gm} \; ,
\qquad b = -\; \frac{\sqrt{\pi}}{12\sqrt{\gm}} \left [
1 + \frac{2\gm s_0^3}{\pi^5\sgm_0} \left ( 28 s_0^3 + \pi^2 -
48 \pi^2 a \right ) \right ] \; .
\ee
Finally, we obtain the asymptotic temperature expansions for the 
condensate fraction
\be
\label{56}
n_0 \simeq 1 \; - \; \frac{s_0^3}{3\pi^2}\; - \;
\frac{\pi^2+12as_0^3}{12\pi^2 s_0}\; t^2
\ee
and for the superfluid fraction
\be
\label{57}
n_s \simeq 1\; - \; \frac{2\pi^2}{45s_0^5}\; t^4 \; .
\ee
Equations (56) and (57) show that $n_s>n_0$.

In order to specify the behavior of $n_0$ and $n_s$ as functions 
of the interaction strength, let us consider two limiting cases, of 
weak and strong interactions. When the interaction is weak, such that 
$\gm\ra 0$, Eqs. (52) and (53) give
\be
\label{58}
s_0 \simeq 2\sqrt{\pi}\; \gm^{1/2} + \frac{16}{3}\; \gm^2 \; ,
\qquad a \simeq -\; \frac{1}{12} \qquad (\gm\ra 0 ) \; .
\ee
Then the sound velocity (51) is
\be
\label{59}
s \simeq 2\sqrt{\pi}\; \gm^{1/2} \; - \; \frac{t^2}{12} \; .
\ee
Remembering condition (46), we see that the considered expansions are 
valid for the low temperatures for which
\be
\label{60}
 t \ll \gm \ll 1 \; .
\ee
For the anomalous average (54), we get
\be
\label{61}
\sgm \simeq \frac{8\gm^{3/2}}{\sqrt{\pi}} \left ( 1 \; - \;
\frac{t^2}{192\gm^2} \right ) \; .
\ee
The condensate fraction (56) becomes
\be
\label{62}
n_0 \simeq 1 \; - \; \frac{8\gm^{3/2}}{3\sqrt{\pi}} \left (
1 + \frac{t^2}{64\gm^2} \right ) \; ,
\ee
which is in agreement with the known temperature expansion for the 
weakly interacting Bose gas [5], first derived by Lee and Yang [34]. 
And for the superfluid fraction (57), we have
\be
\label{63}
n_s \simeq 1\; - \; \frac{\gm^{3/2}}{720\sqrt{\pi}} \left (
\frac{t}{\gm}\right )^4 \; .
\ee
Noting that
\be
\label{64}
\frac{n_s}{n_0} \simeq 1 + \frac{8\gm^{3/2}}{3\sqrt{\pi}}
\left ( 1 + \frac{t^2}{64\gm^2} \right ) \; ,
\ee
we clearly see that $n_s>n_0$, though they are close to each other 
when $\gm\ll 1$ and $t\ll 1$.

In the opposite case of very strong interactions, when $\gm\ra\infty$, 
Eqs. (52) and (53) yield
\be
\label{65}
s_0 \simeq \left ( 3\pi^2\right )^{1/3} - \;
\frac{\pi}{64} \left ( \frac{\pi}{3}\right )^{2/3} \gm^{-3} \; ,
\qquad a \simeq -\; \frac{1}{36} \qquad (\gm\ra\infty) \; .
\ee
The sound velocity (51) behaves as
\be
\label{66}
s \simeq \left ( 3\pi^2\right )^{1/3} -\; \frac{t^2}{36} \; .
\ee
For the anomalous average (54), we find
\be
\label{67}
\sgm \simeq \frac{(9\pi)^{1/3}}{4} \left ( \frac{1}{\gm} \; -\;
\frac{t^2}{9\pi}\right ) \; .
\ee
In this way, we obtain the condensate fraction
\be
\label{68}
n_0 \simeq \frac{\pi}{64}\; \gm^{-3} + O(t^4)
\ee
and the superfluid fraction
\be
\label{69}
n_s \simeq 1 \; - \; \frac{2}{135\pi(9\pi)^{1/3}} \left ( 1 +
\frac{5\pi}{192\gm^3}\right ) t^4 \; .
\ee
The ratio
\be
\label{70}
\frac{n_s}{n_0} \simeq \frac{64}{\pi} \; \gm^3 \qquad 
(\gm\ra\infty)
\ee
shows that $n_s$ is much larger than $n_0$.

\vskip 5mm

{\large{\bf B. Critical Region}}

\vskip 3mm

The analysis of Eqs. (42) to (45) demonstrates that there is the 
critical temperature
\be
\label{71}
t_c = \frac{2\pi}{[\zeta(3/2)]^{2/3}} = 3.312498 \; ,
\ee
where $n_0$, $n_s$, $\sgm$, and $s$ all tend to zero. Temperature 
(71) is the same as the condensation temperature of the ideal Bose 
gas, as it should be in the case of a mean-field approximation for 
atoms with local interactions [5].

Considering the critical region close to $t_c$, when
\be
\label{72}
\frac{s^2}{t_c} \ll 1 \; ,
\ee
we find from Eqs. (43), (44), and (45) the normal fraction
\be
\label{73}
n_1 \simeq \left ( \frac{t}{t_c} \right )^{3/2} +
\frac{s^3}{3\pi^2} \; ,
\ee
the anomalous average
\be
\label{74}
\sgm \simeq \frac{2s^2}{\pi^{3/2}} \; \sqrt{\gm n_0} \; - \;
\frac{st}{2\pi} \; ,
\ee
and the superfluid fraction
\be
\label{75}
n_s \simeq 1  -  \left ( \frac{t}{t_c}\right )^{3/2} +
\frac{\zeta(1/2)}{\zeta(3/2)} \left ( \frac{t}{t_c} 
\right )^{1/2} \frac{s^2}{t_c} \; .
\ee
Calculating the last term in Eq. (75), the dimensional regularization 
was employed.

When temperature $t$ tends to $t_c$, then it is convenient to introduce 
the relative temperature
\be
\label{76}
\tau \equiv 1 \; - \; \frac{t}{t_c} \; \ra \; +0 \; ,
\ee
which tends to zero. Then, solving Eq. (42) results in
\be
\label{77}
s \simeq \sqrt{6\pi\gm}\; \tau^{1/2}
\ee
for any $\gm>0$. Equations (73) and (74) yield the expansions
$$
n_1 \simeq 1 \; - \; \frac{3}{2}\;\tau + 2\; 
\sqrt{\frac{6}{\pi}\; \gm^3}\; \tau^{3/2} \; ,
$$
$$
\sgm \simeq -\; \sqrt{\frac{3\gm}{2\pi} } \; t_c \tau^{1/2} +
\frac{12\gm}{\sqrt{\pi}}\; \tau + \sqrt{\frac{3\gm}{2\pi} } \;
t_c \tau^{3/2} \; .
$$
Using these, we obtain the condensate fraction
\be
\label{78}
n_0 \simeq \frac{3}{2}\; \tau - 2\; 
\sqrt{\frac{6}{\pi}\; \gm^3} \; \tau^{3/2}
\ee
and the superfluid fraction
\be
\label{79}
n_s \simeq \frac{3}{2} \left [ 1 + 
4\pi \frac{\zeta(1/2)\gm}{\zeta(3/2)t_c} \right ] \tau \; .
\ee
As we see, because of the ratio
\be
\label{80}
\frac{n_s}{n_0} \simeq 1 + 
4\pi \frac{\zeta(1/2)\gm}{\zeta(3/2)t_c} \; ,
\ee
the superfluid fraction is again larger than the condensate 
fraction for all $\gm>0$.

Strictly speaking, the above expansions for the critical region are
valid for not too large gas parameter $\gm$. This is because the first
term in Eq. (29) has been obtained using the dimensional regularization, 
which results in Eq. (30). The dimensional regularization is known [5] 
to be asymptotically exact in the limit of weak interactions. The 
analytical continuation of Eq. (30) to finite interactions may become not 
accurate for large values of the latter. In the above expansions, we 
constantly meet the combination $\gm t$, which is to be smaller than one 
to make the expansions quantitatively correct. From the inequality $\gm t 
\leq 1$, in the critical region, when $t\approx t_c$, we get $\gm \leq 
0.3$. Considering in what follows the critical properties of the system, 
for $\gm > 0.4$, we keep in mind that the behaviour of thermodynamic 
characteristics for these values of $\gm$ is only approximate.

The asymptotic behavior of $n_0$ and $n_s$ in expansions (78) and 
(79) makes it apparent that both $n_0$ and $n_s$ tend simultaneously 
to zero at $t_c$, so that BEC coincides with the superfluid transition. 
This phase transition is of {\it second order}, which agrees with the 
universality theory. According to the latter, the considered Bose 
system with BEC belongs to the universality class of the 3-dimensional 
$O(2)$-symmetric spin model, hence, must display the second-order phase 
transition [5,35,36]. It is worth noting the importance of the anomalous 
average $\sgm$. Though it tends to zero, as $t\ra t_c$, but it cannot 
be neglected, since $n_0$ also tends to zero. Neglecting the anomalous 
average $\sgm$ would result in the first-order phase transition [5], 
which is not correct. In addition, omitting $\sgm$ would make the system 
unstable at all temperatures [31].

\vskip 5mm

{\large{\bf C. Numerical Solution}}

\vskip 3mm

In order to investigate the behavior of the characteristic quantities 
in the whole range of temperatures $t\in[0,t_c]$ and for arbitrary 
interactions $\gm\geq 0$, we solve numerically the system of equations 
(42), (43), (44), and (45), together with the relation $n_0=1-n_1$. 
Figure 1 presents the normal fraction (43) as a function of the 
dimensionless temperature (39) and the interaction strength (38). 
In Fig. 2, we show the anomalous average (44) as a function of the 
same variables $t$ and $\gm$. The condensate fraction is depicted 
in Fig. 3 and the superfluid fraction Fig. 4. Increasing temperature 
depletes both $n_0$ and $n_s$. But their dependence on the interaction 
strength $\gm$ is not the same. Increasing $\gm$ always strengthens 
superfluidity, and $n_s$ becomes larger. However the dependence of 
$n_0$ on $\gm$ is not trivial. At zero temperature, the condensate 
fraction diminishes with increasing $\gm$. But at finite temperatures, 
$n_0$, first, increases with rising $\gm$ and then diminishes. This 
nonmonotonic behavior of $n_0$ is connected with the nonmonotonic 
dependence of the anomalous average $\sgm$ on $\gm$. The dimensionless 
sound velocity (40), given by the solution of Eq. (42), is displayed 
in Fig. 5. Temperature always diminishes $s$, while interactions make 
$s$ larger.

\section{Discussion}

We have presented a detailed analysis of the condensate, $n_0$, and 
superfluid, $n_s$, fractions for arbitrary interactions strengths and 
for all temperatures in the internal $[0,T_c]$. The consideration is 
based on the self-consistent mean-field theory [20--25]. For the limiting 
cases of low temperatures and for those in the critical region, we derive 
analytic expressions for $n_0$ and $n_s$. And for the whole range of 
temperatures and interactions, we accomplish numerical calculations. The 
appearance of BEC and superfluidity occurs simultaneously at the critical 
temperature $T_c$. This transition is of second order, as it must be 
according to the universality theory. The superfluid fraction is practically 
always larger than the condensate one. They coincide only at $T_c$ or for the 
case of the ideal Bose case.

It is important to stress the crucial role of the anomalous average (44). 
If it would be neglected, the phase transition would be of first order, 
which is not correct. And, moreover, neglecting the anomalous average 
renders the consideration not self-consistent and the system unstable.

We may mention that the relation between the condensate and superfluid 
fractions can be connected with the infrared behavior of the 
single-particle Green function $G_{11}(\bk,\om)$ in the momentum-energy 
representation. This asymptotic behavior, having the form
$$
\left | G_{11}(\bk,0) \right | \simeq \frac{mn_0}{n_sk^2} \qquad
(k\ra 0) \; ,
$$
has been, first, obtained by Bogolubov, analyzed in great detail in 
Refs. [37], and summarized in his books [29,30]. As is evident, the 
Bogolubov infrared expression for $G_{11}(\bk,0)$ can be rewritten 
as the limit
$$
\frac{n_0}{n_s} = \lim_{k\ra 0} \; \frac{k^2}{m} \left |
G_{11}(\bk,0) \right | \; .
$$
The same asymptotic infrared expression has also been rederived 
and discused in Refs. [38--40]. Actually, the above relation is not 
an explicit definition of the ratio $n_0/n_s$ but it is an implicit 
equation, since $G_{11}$ itself is a complicated function of $n_0$. 
In our paper, the relation between $n_0$ and $n_s$ is given by 
$n_0=1-n_1$, with $n_1$ defined in Eq. (43), and by Eq. (45) for 
$n_s$. These equations (43) and (45) also are not the explicit 
expressions for $n_0$ and $n_s$, but are a part of the system of 
equations (42), (43), (44), and (45). Solving these equations makes 
it possible to extract the condensate and superfluid fractions as 
functions of temperature and interaction strength.

There is no simple general relation between the condensate and superfluid 
fractions, represented as explicit functions of temperature and gas parameter. 
In some limiting cases, it is possible to find asymptotic relations between 
these functions. For example, at low temperature, the fractions are related 
through Eq. (50). When, in addition, the interaction is asymptotically weak, 
satisfying Eq. (60), then both these fractions are close to each other, as 
in Eqs. (62), (63), and (64).

In the opposite case, when temperatures are low, but the interaction is 
strong, the condensate fraction is drastically depleted, in agreement with 
Eq. (68). At the same time, the superfluid fraction, according to Eq. (69), 
is close to one. The limiting case, corresponding to Eq. (70), could be 
realized in low-temperature experiments with cold atoms by increasing their 
scattering length by means of the Feshbach-resonance techniques. Then the 
state of the Bose gas can be achieved, which contains a very small condensate 
fraction, though being almost completely superfluid. 

Another way of creating the state with a tiny condensate fraction, but with 
a large superfluid fraction, close to one, could be by loading atoms into 
an optical lattice, so that to reduce their effective mass. Diminishing the 
effective mass, as follows from Eq. (7), is equivalent to the increase of 
the effective interaction. In the one-dimensional case, the strengthening 
of atomic interactions would result in the effective fermionization of bosons, 
corresponding to the Girardeau mapping (see review article [7]).

The analysis of the present paper can be straightforwardly extended to 
nonuniform systems, such as atomic systems in trapping potentials. For 
shallow traps, the local-density approximation (see Refs. [2.3]) can be 
employed. Then the overall consideration remains practically the same as 
above, merely with the appearing dependence on the real-space  variable. In 
that case, the equations of the present paper can be interpreted as being 
written for the center of the trap.

In the general case of an arbitrary trapping potential, we again could 
follow the same steps as above, just with some technical complications, 
caused by the system nonuniformity [20-22, 32]. Then the main difference 
is that the Hartree-Fock-Bogolubov decoupling for the field operators of 
uncondensed atoms should be done in the real-space  representation and the 
general form of the Bogolubov canonical transformations [29, 30] has to be 
used, as in Ref. [32]. Following this way for atoms in a trap with a trapping 
potential $U(\br)$ and local interactions $\Phi(\br)=\Phi_0\dlt(\br)$, we can 
use the Hartree-Fock-Bogolubov approximation for a nonuniform matter and the 
canonical transformations as in Ref. [32]. Then, introducing the notation
$$
\om(\br) = -\; \frac{\nabla^2}{2m} + U(\br) + 2\Phi_0 \rho(\br) -\mu_1 \; ,
$$
$$
\Dlt(\br) = \Phi_0\left [ \rho_0(\br) +\sgm_1(\br) \right ] \; ,
$$
where the total density of atoms is the sum of the density of condensed and 
uncondensed atoms,
$$
\rho(\br)=\rho_0(\br) + \rho_1(\br) \; ,
$$
we obtain the Bogolubov equations, providing the diagonalization of the 
Hamiltonian, 
$$
\om(\br) u_k(\br) +\Dlt(\br)v_k(\br) = \ep_k u_k(\br) \; , \qquad
\om^*(\br) v_k(\br) +\Dlt^*(\br)u_k(\br) = -\ep_k v_k(\br) \; .
$$
Here $k$ is a multi-index labelling the solutions to the above eigenproblem.
Since $\om(\br)$ is real, we can take $\Dlt(\br)$ to be also real. The 
Bogolubov equations define the spectrum of collective excitations and the 
coefficient functions $u_k(\br)$ and $v_k(\br)$. The latter functions have 
to satisfy the canonical conditions, similar to the  uniform case, which 
would guarantee the Bose commutation relations of the field operators. Note 
that the anomalous average in these equations cannot be omitted, since, as 
has been  shown above, the anomalous average is of the order of the normal 
average. Omitting the anomalous average would lead to incorrect results.
    
The space-dependent condensate density
$$
\rho_0(\br) = |\eta(\br)|^2 \; 
$$
is defined through the condensate wave function $\eta(\br)$, satisfying the 
equation
$$
\left [ -\; \frac{\nabla^2}{2m} + U(\br)\right ] \eta(\br) +
\Phi_0 \left [ |\eta(\br)|^2 \eta(\br) + 2 \rho_1(\br)\eta(\br) +
\sgm_1(\br) \eta^*(\br) \right ] = \mu_0 \eta(\br) \; .
$$

If $\Dlt(\br)$ is real, then $\sgm_1(\br)$ is also real. For an equilibrium 
system, the condensate function $\eta(\br)$ is real. The parameter $\mu_0$ is 
given by the  normalization condition
$$
N_0 = \int |\eta(\br)|^2 \; d\br \; .
$$
The normal and anomalous averages, $\rho_1(\br)$ and $\sgm_1(\br)$, are defined 
in the  standard way. The number of uncondensed atoms is
$$
N_1 = \int \rho_1(\br)\; d\br \; .
$$
The condition of the condensate existence
$$
\min_k \ep_k = 0 \; , \qquad \ep_k \geq 0 \; ,
$$
gives $\mu_1=\mu_1(T)$ as a function of temperature and other system 
parameters, such as the gas parameter. Hence, the number of uncondensed 
atoms, $N_1=N_1(T)$, is also a function of temperature and all other 
system parameters, As as result, we can find the number of condensed 
atoms $N_0=N_0(T)$ as $N_0(T)=N-N_1(T)$ and, respectively, the condensate 
fraction $N_0/N$.

Thus, the whole procedure for trapped atoms is ideologically the same as 
for the uniform system. The technical complications come from the necessity 
to solve the equation for the condensate function and the Bogolubov equations 
for the coefficient functions and  the collective spectrum. This, generally, 
requires the usage of extensive numerical  calculations.

\newpage

\begin{center}
{\Large{\bf Figure Captions}}
\end{center}

\vskip 1cm

{\bf Fig. 1}. Fraction of uncondensed atoms $n_1=n_1(t,\gm)$ 
as a function of the dimensionless temperature $t$ and of the 
interaction strength $\gm$.

\vskip 5mm

{\bf Fig. 2}. Anomalous average $\sgm=\sgm(t,\gm)$ as a function 
of the variables $t$ and $\gm$.

\vskip 5mm

{\bf Fig. 3}. Condensate fraction $n_0=n_0(t,\gm)$ as a function 
of the variables $t$ and $\gm$.

\vskip 5mm

{\bf Fig. 4}. Superfluid fraction $n_s=n_s(t,\gm)$ as a function 
of the variables $t$ and $\gm$.

\vskip 5mm

{\bf Fig. 5}. Dimensionless sound velocity $s=s(t,\gm)$ as a function 
of temperature $t$ and the interaction strength $\gm$.

\newpage

\begin{figure}[h]
\centerline{\psfig{file=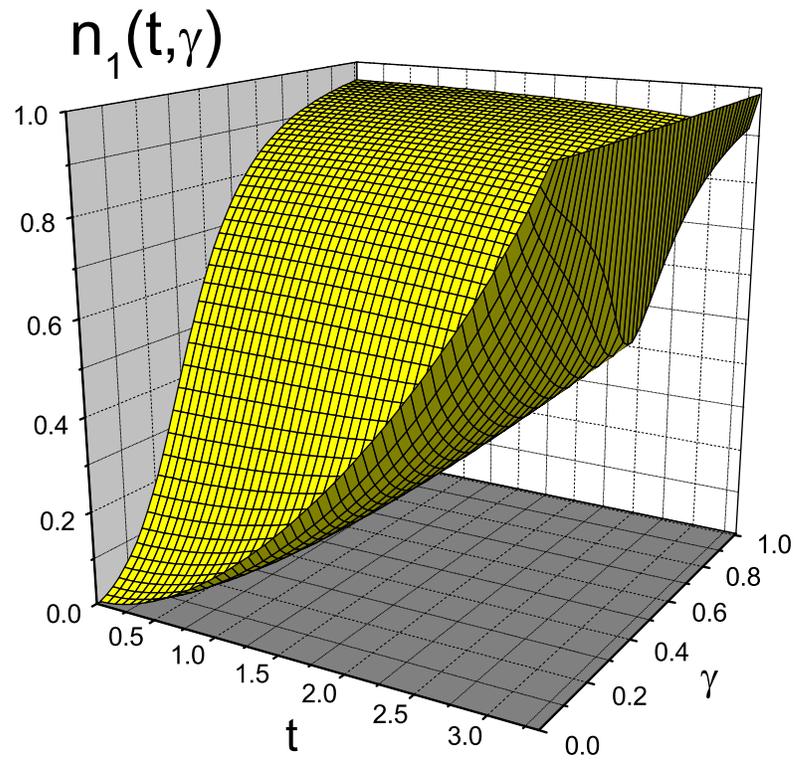,angle=0,width=15cm}}
\vskip 1cm
\caption{Fraction of uncondensed atoms $n_1=n_1(t,\gm)$
as a function of the dimensionless temperature $t$ and of the
interaction strength $\gm$.}
\label{fig:Fig.1}
\end{figure}

\newpage

\begin{figure}[h]
\centerline{\psfig{file=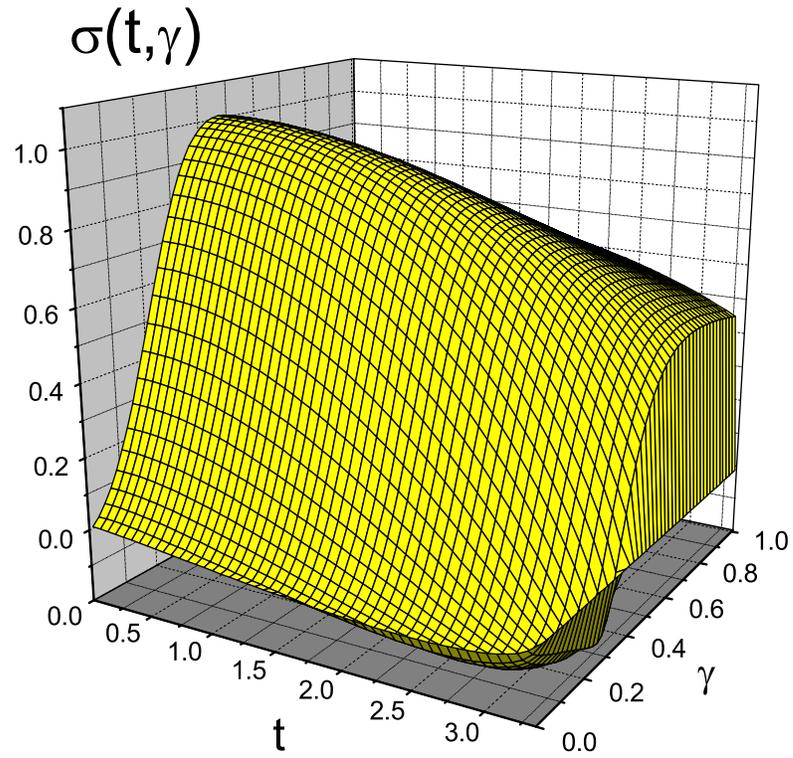,angle=0,width=15cm}}
\vskip 1cm
\caption{Anomalous average $\sgm=\sgm(t,\gm)$ as a function
of the variables $t$ and $\gm$.}
\label{fig:Fig.2}
\end{figure}

\newpage

\begin{figure}[h]
\centerline{\psfig{file=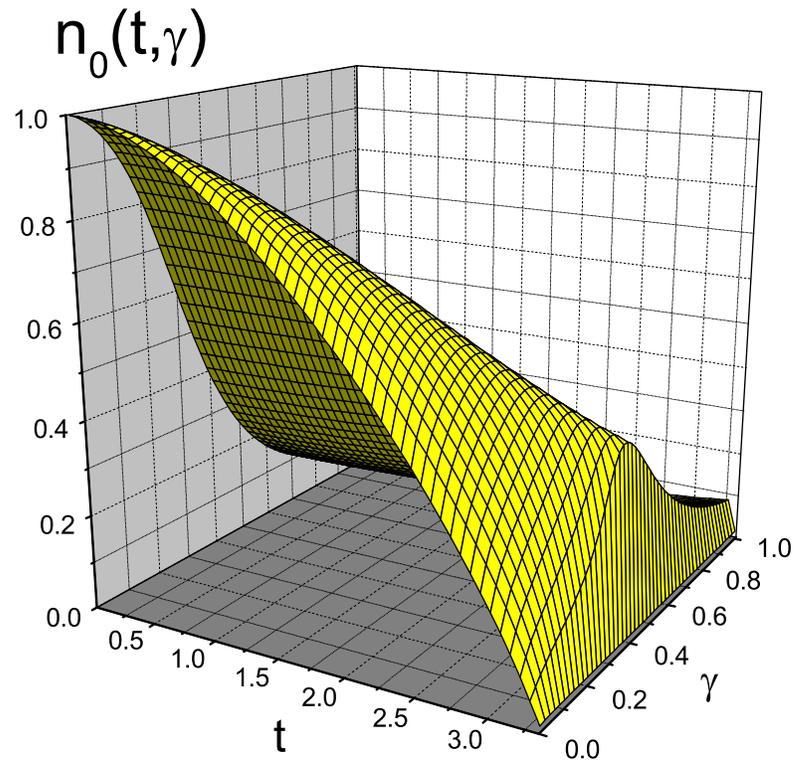,angle=0,width=15cm}}
\vskip 1cm
\caption{Condensate fraction $n_0=n_0(t,\gm)$ as a function
of the variables $t$ and $\gm$.}
\label{fig:Fig.3}
\end{figure}

\newpage

\begin{figure}[h]
\centerline{\psfig{file=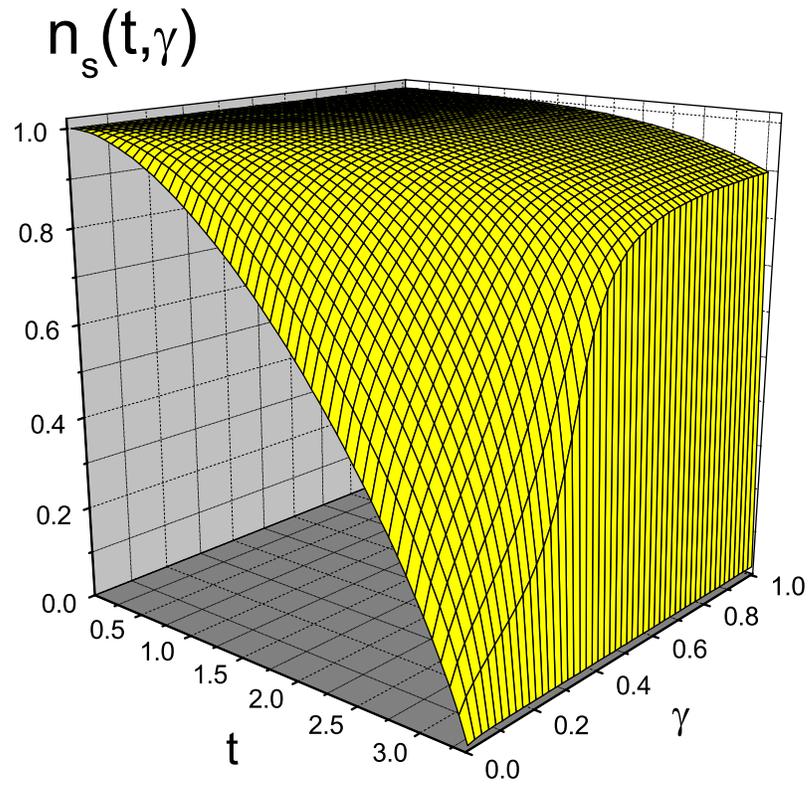,angle=0,width=15cm}}
\vskip 1cm
\caption{Superfluid fraction $n_s=n_s(t,\gm)$ as a function
of the variables $t$ and $\gm$.}
\label{fig:Fig.4}
\end{figure}

\newpage

\begin{figure}[h]
\centerline{\psfig{file=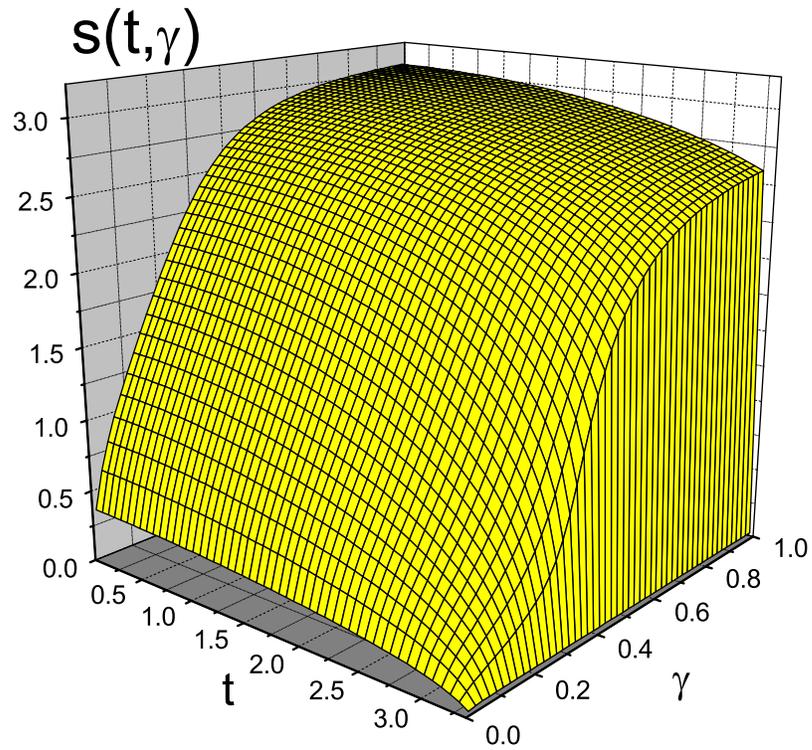,angle=0,width=15cm}}
\vskip 1cm
\caption{Dimensionless sound velocity $s=s(t,\gm)$ as a function
of temperature $t$ and the interaction strength $\gm$.}
\label{fig:Fig.5}
\end{figure}

\end{document}